\documentclass[aps,prb,floatfix,twocolumn,showpacs]{revtex4}
\usepackage{epsfig}
\usepackage{verbatim}	 
\usepackage{amssymb}
\usepackage{bbold}
\usepackage{amsmath}
\usepackage{psfrag}
\usepackage{color}
\usepackage{ulem}
\newcommand{\bi}{\bibitem}
\newcommand{\eps}{\varepsilon}
\newcommand{\GL}{\Gamma_{\rm L}}
\newcommand{\GR}{\Gamma_{\rm R}}
\newcommand{\TL}{T_{\rm L}}
\newcommand{\TR}{T_{\rm R}}
\newcommand{\UL}{U_{\rm L}}
\newcommand{\UR}{U_{\rm R}}

\newcommand{\df}{(-\partial_\eps f)}
\newcommand{\COP}{{\rm COP}}
\begin{document}

\hsize\textwidth\columnwidth\hsize\csname@twocolumnfalse\endcsname

\title{Scattering Theory of Nonlinear Thermoelectricity in Quantum Coherent Conductors}

\author{Jonathan Meair$^1$ and Philippe Jacquod$^{1,2}$}
\affiliation{$^1$Physics Department, University of Arizona, Tucson, AZ 85721, USA \\$^2$College of Optical Sciences, University of Arizona, 
Tucson, AZ 85721, USA}

\vskip1.5truecm
\begin{abstract}
We construct a scattering theory of weakly nonlinear thermoelectric transport through
sub-micron scale conductors. The theory incorporates the leading 
nonlinear contributions in temperature and voltage biases to the charge and heat currents. 
Because of the finite capacitances of sub-micron scale conducting circuits, fundamental conservation
laws such as gauge invariance and current conservation require special care to be preserved. 
We do this by extending the approach of Christen and B\"uttiker [Europhys. Lett. {\bf 35}, 523 (1996)]
to coupled charge and heat transport. In this way we write relations 
connecting nonlinear transport coefficients in a  manner
similar to Mott's relation between the linear thermopower and the linear conductance.
We derive sum rules that nonlinear transport coefficients
must satisfy to preserve gauge invariance and current conservation. We illustrate our theory
by calculating the efficiency of heat engines and the coefficient of performance of 
thermoelectric refrigerators
based on quantum point contacts and
resonant tunneling barriers. We identify in particular rectification effects that increase device performance.
\end{abstract}

\maketitle

\section{Introduction}

Today's energy challenges are calling for novel technologies to produce, store and use energy~\cite{Chu12}.
These challenges have among others renewed the interest in the old problem of 
thermoelectricity~\cite{Mah97}. Research in thermoelectricity is predominantly 
directed towards finding
new materials with better thermoelectric properties~\cite{Maj04,Sny08,Sha11}, however a recently proposed
alternative approach has been to try and tailor low-dimensional micro- and nano-structured
materials~\cite{Dres07}, which may exhibit the sharp oscillations in the electronic
density of states necessary to trigger large thermoelectric effects~\cite{Mah96}.
With the ongoing trend 
towards circuit miniaturization in electronics, heat evacuation becomes 
a major problem
and it is accordingly important to explore how heat currents can be controlled at
sub-micron scales. Thus, there is ample motivation for investigating
thermoelectricity in micro- and nanodevices~\cite{Gia06}. 
To name but few examples of recent investigations, 
sub-micron electrically controlled heat rectifiers~\cite{Sch08}, thermally driven 
electric rectifiers~\cite{Sot12,Mat12},
thermoelectric refrigerators~\cite{Edw95,Pra09,Rob12.1} and
heat converters~\cite{Hum02,San11,Mur12}  have been proposed,
and thermoelectric effects 
in molecular transport have been investigated~\cite{Red07,Bah08,Berg09,Noz10,Ent10,Yee11,Nik12}.
On a more fundamental level, thermoelectric effects provide information on the nature
of the underlying quantum fluid, for instance, breakdowns of the Wiedemann-Franz law~\cite{Ashcroft}
may reflect non Fermi liquid behaviors~\cite{hill01,Col05,Pod07}, quantum 
interference effects~\cite{Vav05,Bal12}
or electric and heat transport mediated by both electron and hole quasiparticles in 
Andreev systems~\cite{Jian05,Vir07,RobTP}.

The field of thermoelectricity has 
until now overlooked the fact that 
thermoelectric devices often operate in the nonlinear regime of transport (See for instance Refs.~[\onlinecite{Red07,Dzu93,Sho01}]). 
Rectification effects
occur  in that regime~\cite{Cas02,San04,Spi04,And06,Mea12}, and their influence on thermoelectricity is 
not yet understood. There are only few works we know about which have
discussed nonlinear thermoelectricity. 
Ref.~[\onlinecite{Gri91}] calculated the thermopower to quadratic order in temperature gradients and
Refs.~[\onlinecite{Kul94,Free06,Zeb07}] focused on the nonlinear Peltier coefficient. These works 
considered bulk systems with large capacitances, 
where accordingly the build-up of local electrostatic potentials can be neglected. 
This neglect is no longer justified in small constrictions, where such potentials have to be
self-consistently connected to the applied biases in order to preserve gauge 
invariance~\cite{Chri96}. Refs.~[\onlinecite{Bog99,Cip04}] discussed nonlinear thermoelectricity in
such constrictions but did not introduce local potentials, thereby taking the risk of violating gauge invariance. 
Refs.~[\onlinecite{Dzu93,Rob12.1,San12}] introduced self-consistently determined 
homogeneous Hartree potentials to preserve gauge invariance. 

In this manuscript we generalize the  scattering 
theory of weakly nonlinear transport of Ref.~[\onlinecite{Chri96}] to coupled electric and heat transport.
We go beyond Ref.~[\onlinecite{San12}] which considered only electric currents in the presence of 
voltage and temperature biases, by additionally treating heat currents. One of our motivations
is to provide criteria for enhanced thermoelectric efficiency in the nonlinear regime.
The standardly used figure of merit $ZT$  is constructed out of linear thermoelectric
coefficients only~\cite{Mah97}, thus while any measure 
of thermodynamic efficiency in the linear regime can only be a function of $ZT$, 
it is expectable that 
$ZT$ loses its predictive power in the nonlinear regime. It is  often the case, and we illustrate
this below, that the maximal efficiency predicted by $ZT$ is attained 
outside the linear regime.
In this article we accordingly focus our attention on
thermodynamic efficiencies and coefficients of performance as indicators of thermoelectric
performance. 
In particular we investigate how they behave 
upon increasing sources of nonlinearities such as temperature and voltage
biases and capacitive couplings. A key aspect of our theory is that in the nonlinear
regime of transport, the state of the system is self-consistently determined by the conditions
under which transport proceeds. We will take this self-consistency into account by introducing
piecewise constant local electrostatic potentials. We will go beyond
Refs.~[\onlinecite{Dzu93,Rob12.1,San12}] and consider more than one such potential.
In the scattering theory, 
gauge invariance and current conservation are naturally preserved in
the linear regime by the unitarity of the scattering matrix. In the nonlinear regime, 
they are reflected by sum rules connecting nonlinear transport coefficients. Such sum rules
were derived in Ref.~[\onlinecite{Chri96}] for purely electric transport and below we give a complete
list of sum rules ensuring gauge invariance and current conservation for weakly nonlinear 
thermoelectric transport. 

The paper is organized as follows. In Section~\ref{scat}, we construct a scattering theory of weakly 
nonlinear thermoelectric transport valid to quadratic order in temperature and voltage biases.
We express linear and nonlinear transport coefficients in terms of the scattering matrix and the 
local potentials generated in the system by the biases. In Section~\ref{locpot}, we discuss these local
potentials and recall how they can be determined in practice. In Section~\ref{sumrules} we write down
sum rules that have to be satisfied for the theory to preserve charge and current conservation as well
as gauge invariance. In Section~\ref{efficiency} we briefly 
discuss performance metrics for 
thermoelectric devices. In Section~\ref{illustration} 
we apply our theory to two particular devices, a quantum point contact and a resonant tunneling 
double-barrier. We find in particular that the bias range of operation of these devices is such that 
our weakly nonlinear approximation gives a rather accurate description of the exact thermodynamic efficiency. 
Conclusions are presented in Section~\ref{conclusions}.


\section{Scattering Approach to Thermoelectricity}\label{scat}

\begin{figure}[ht]
\psfrag{Vg}[][][1.5]{$V_g$}
\psfrag{U}[][cl][1.5]{$U({\vec r})$}
\psfrag{m1}{$\mu_1$}
\psfrag{m2}{$\mu_2$}
\psfrag{m3}{$\mu_3$}
\psfrag{T1}{$T_1$}
\psfrag{T2}{$T_2$}
\psfrag{T3}{$T_3$}
\psfrag{I1}{$I_1,J_1$}
\psfrag{I2}{$I_2,J_2$}
\psfrag{I3}{$I_3,J_3$}
\includegraphics[width=3in]{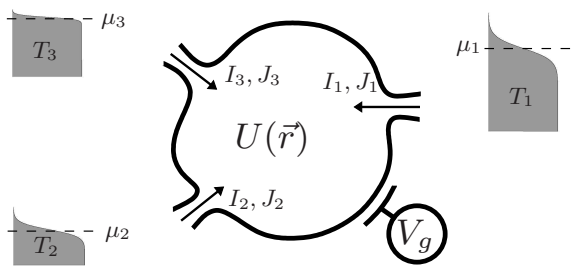}
\caption{Sketch of a multi-terminal coherent conductor capacitively coupled to an external gate and connected 
to three external Fermi liquid reservoirs. The electronic Fermi distributions of the reservoirs are depicted. In the nonlinear regime, 
an inhomogeneous electrostatic potential 
$U({\vec r})$ develops inside the conductor, as it is
induced by the applied voltages and temperatures at the reservoirs.
The arrows indicate our convention that electric and heat currents are positive when they
flow into the conductor.}\label{fig:generic}
\end{figure}

In linear response, the scattering theory to thermoelectric transport was worked out
by Butcher~\cite{But90}. We follow his approach and consider heat and electric transport mediated
only by electron quasiparticles and extend it to the nonlinear regime.
Considering a coherent conductor connected to $i=1,2,...M$ external
reservoirs via reflectionless leads (see Fig.~\ref{fig:generic}), 
the electric ($I_i$)  and heat ($J_i$) currents through contact $i$ are given by
\begin{subequations}\label{eq:currents_initial}
\begin{align}
I_i &= \frac{2e}{h}\sum_j \int d\eps \;f_j(\eps)\; A_{ij}\left[\eps,U({\vec r})\right] \, , \\
J_i &= \frac{2}{h}\sum_j \int d\eps \;(\eps-eV_i)\;f_j(\eps) \; A_{ij}\left[\eps,U({\vec r})\right] \, .\label{eq:1b}
\end{align}
\end{subequations}
We took the convention that both electric and heat currents are positive when they
enter the conductor from a reservoir. Thus, for example, the conductor is cooled if $\sum_i J_i < 0$
and heated otherwise.
In Eqs.~(\ref{eq:currents_initial}), 
$\eps \equiv E - E_F$ is the energy of an electron counted from the Fermi energy,
$f_j(\eps)=\left\{1+\text{exp}\left[(\eps-eV_j)/k_B T_j\right]\right\}^{-1}$ is the Fermi function in 
reservoir $j$, maintained at a voltage $V_j$ and a temperature $T_j$,
$A_{ij}\left[\eps,U({\vec r})\right] = N_i[\eps,U({\vec r})] \delta_{i,j} - \mathcal{T}_{ij}\left[\eps,U({\vec r})\right]$ is given by the number $N_i$ of transport channel in lead $i$ and the scattering matrix ${\bf s}$
via $\mathcal{T}_{ij} = {\rm Tr}[{\bf s}^\dagger_{ij}{\bf s}_{ij}]$ where ${\bf s}_{ij}$ is the sub-block of ${\bf s}$
connecting terminals $j$ to $i$ and
$U({\vec r})$ is the local potential in the conductor. 
In this manuscript, we take $N_i$ to be independent of energy and biases. 
The electrostatic potential induced within the conductor (see Fig.~\ref{fig:generic}) is itself dependent on the voltages and temperatures in the reservoirs, i.e., $U({\vec r},\{V_k\},\{T_k\})$. Expanding the currents 
in Eqs.~(\ref{eq:currents_initial}) to quadratic order in temperature differences, $\theta_j\equiv T_j - T_0$, and voltage biases one obtains
\begin{subequations}\label{eq:weaknlin}
\begin{eqnarray}
I_i &=& \sum_{j} \left(G_{ij} V_j  + B_{ij} \theta_j\right) \nonumber\\
&&+ \sum_{j,k} \left[ G_{ijk} V_j V_k + B_{ijk} 	\theta_j \theta_k + Y_{ijk} V_j \theta_k \right] \, , \\
J_i &=& \sum_{j} \left(\Gamma_{ij} V_j  + \Xi_{ij} 	\theta_j\right) \nonumber\\
&&+ \sum_{j,k} \left[\Gamma_{ijk} V_j V_k + \Xi_{ijk} 	\theta_j \theta_k + \Psi_{ijk} V_j \theta_k \right].
\end{eqnarray}
\end{subequations}
The linear response coefficients were calculated in Ref.~[\onlinecite{But90}]. They 
are given by
\begin{subequations}
\begin{eqnarray}
G_{ij} &=&  \frac{2e^2}{h} \, \int d\eps \; \df \; A_{ij}(\eps) \, , \\
B_{ij} &=& \frac{2e}{hT_0} \int d\eps \;\df \;\eps \; A_{ij}(\eps) \, , \label{onsag1}\\
\Gamma_{ij} &=& \frac{2e}{h} \;\;\, \int d\eps \;\df \;\eps \; A_{ij}(\eps) \, , \label{onsag2}\\
\Xi_{ij} &=& \frac{2}{h T_0} \int d\eps  \;\df\;\eps^2\; A_{ij}(\eps) \, ,
\end{eqnarray}
\end{subequations}
where $f(\eps)=\left\{1+\text{exp}\left[\eps/k_B T_0\right]\right\}^{-1}$.
The temperature $T_0 \ne 0$ is in principle arbitrary, however, better convergence
of the expansion of Eqs.~(\ref{eq:weaknlin}) is obtained when $T_0$ is taken 
as the temperature in one of the reservoirs or the
average of the reservoir temperatures. In particular, the sum rules given below in Section~\ref{sumrules}
guarantee that all currents $I_i$ and $J_i$ vanish at equilibrium, regardless of the choice of $T_0$.
Eqs.~(\ref{onsag1}) and (\ref{onsag2}) show that the
reciprocity relation $\Gamma_{ij} = B_{ij} T_0 $ is automatically
satisfied (see Ref.~[\onlinecite{Onsager}] for a detailed discussion of Onsager relations in the 
linear regime).

These linear response coefficients are determined by the equilibrium state of the system for which we set
$U^{(eq)}({\vec r})=0$. However, in the nonlinear regime, we must consider how the biases affect the local potential. We do that by introducing characteristic potentials~\cite{Chri96,San12} 
\begin{align}\label{eq:charpot}
U({\vec r}) = \sum_k \left[ u_k({\vec r}) \;V_k + z_k({\vec r})\; \theta_k \right],
\end{align}
which describe how $U(\vec{r})$ depends on the 
biases to leading order. Truncating the expansion of $U(\vec{r})$ to linear order in the biases is
sufficient to correctly capture currents to quadratic order.
Gauge invariance requires that $\sum_k u_k(\vec{r}) = 1$~\cite{Chri96}, 
however there is no such relation involving $z_k(\vec{r})$.
The nonlinear response coefficients are then
\begin{subequations}\label{eq:nonlincoeff}
\begin{eqnarray}
G_{ijk} &=& \frac{e^3}{h} \int d\eps \;\df \; \alpha_{ijk}(\eps) \, , \\
B_{ijk} &=& \frac{e}{hT_0^2} \int d\eps \;\df \;\eps \; \beta_{ijk}(\eps) \, , \\
Y_{ijk} &=& \frac{2e^2}{hT_0} \int d\eps \;\df \; \gamma_{ijk}(\eps)\, ,  \\
\Gamma_{ijk} &=& \frac{e^2}{h} \int d\eps \;\df \;\eps \; \alpha_{ijk}(\eps) \nonumber \\
&& -G_{ik} \left(\delta_{i,j}-\frac{1}{2} \delta_{j,k} \right) \, , \\
\Xi_{ijk} &=& \frac{1}{h T_0^2} \int d\eps \;\df \;\eps^2 \; \beta_{ijk}(\eps) + \frac{\Xi_{ij}}{2 T_0}\delta_{j,k} \, , \\
\Psi_{ijk} &=& \frac{2e}{h T_0} \int d\eps \;\df \;\eps \; \gamma_{ijk}(\eps) \nonumber \\
&& -B_{ik} \left(\delta_{i,j} - \delta_{j,k}\right),
\end{eqnarray}
\end{subequations}
where we defined 
\begin{subequations}\label{eq:abc}
\begin{eqnarray}
\alpha_{ijk}(\eps) &=& \int d{\bf r} \frac{\delta A_{ij}(\eps)}{e\delta U({\vec r})} \left[2 u_k({\vec r}) - \delta_{j,k}\right] \, , \\
\beta_{ijk}(\eps) &=& \int d{\bf r} \frac{\delta A_{ij}(\eps)}{e\delta U({\vec r})}  \left[2 e T_0 z_k({\vec r}) -\eps \delta_{j,k} \right] \, , \\
\gamma_{ijk}(\eps) &=& \int d{\bf r} \left(\frac{\delta A_{ij}(\eps)}{e\delta U({\vec r})} \left[e T_0 z_k({\vec r}) - \eps \delta_{j,k} \right] \right. \nonumber \\
&& \left. \;\;\;\;\;\;\;\;\;\, + \frac{\delta A_{ik}(\eps)}{e\delta U({\vec r})} \, \eps \, u_j({\vec r})\right).
\end{eqnarray}
\end{subequations}
Several of the nonlinear response coefficients of Eqs.~(\ref{eq:nonlincoeff}) 
differ primarily by a factor of $\eps$ inside the energy integral. 
Using a Sommerfeld expansion, we find that at low temperature the following relations between nonlinear response coefficients must hold,
\begin{subequations}
\begin{align}
\frac{\Gamma_{ijk}+G_{ik}(\delta_{i,j}-\delta_{j,k}/2)}{G_{ijk}} 
&= 
\frac{(\pi k_B T_0)^2}{3e} \; \partial_\eps \text{ln} \, G_{ijk}  \, , \\
\frac{\Psi_{ijk}+B_{ik}(\delta_{i,j}-\delta_{j,k})}{Y_{ijk}} 
&= 
\frac{(\pi k_B T_0)^2}{3e} \; \partial_\eps \text{ln} \, Y_{ijk}  \, , \\
\frac{B_{ijk}}{\Xi_{ijk}-\Xi_{ij}\delta_{j,k}/(2T_0)}
&= 
e  \; \partial_\eps \text{ln} \, \left[\Xi_{ijk}-\frac{\Xi_{ij}\delta_{j,k}}{2T_0}\right] \, ,
\end{align}
\end{subequations}
where all derivatives on the right-hand side are evaluated at the Fermi energy, $\eps=0$.
These relations are the nonlinear counterpart to the Mott relation~\cite{Ashcroft}
$S=-(\pi^2 k_B^2 T_0/3e)\partial_\eps \ln G $ between the thermopower $S=-B_{12}/G_{12}$ 
and the logarithmic
derivative of the conductance $G \equiv G_{12}$ in two-terminal conductors~\cite{thermopower_note}.
There is no counterpart to the  Onsager reciprocity relations for nonlinear coefficients,
because, as can be seen from Eqs.~(\ref{eq:nonlincoeff}), they all have different energy integrands.
The absence of such relations may be 
attributed to the breaking of microscopic reversibility in the nonlinear regime, where local potentials 
are self-consistently related to transport, and where their dependence on temperature [the coefficients 
$z_k$ in Eq.~(\ref{eq:charpot})] is not necessarily related to their dependence on voltages [the coefficients 
$u_k$ in Eq.~(\ref{eq:charpot})]. 

\section{Characteristic potentials}\label{locpot}

Our main task is next to determine the characteristic potentials defined in Eq.~(\ref{eq:charpot}).
We briefly summarize how this can be done~\cite{Chri96,San12}.
We consider first how additional charges are injected into the conductor as it is driven out of equilibrium. Biased reservoirs inject a net addition of charge, called the bare charge, $\delta q^{(b)}$. The presence
of $\delta q^{(b)}$ induces 
an electrostatic potential shift within the conductor which, in its turn,
generates a screening charge $\delta q^{(s)}$. 
The net injected charge is the sum of the two, 
$\delta q({\vec r}) = \delta q^{(b)}({\vec r}) + \delta q^{(s)}({\vec r})$.
To leading order the bare charge can be expressed as a linear function of the biases,
\begin{align}
\delta q^{(b)}({\vec r}) = \sum_j D_j^{V} ({\vec r}) V_j + \sum_j D_j^{\theta} ({\vec r}) \theta_j, \label{bare}
\end{align}
where $D_j^V({\vec r})=e^2\int d\eps \df\; \nu_j^p({\vec r},\eps) $ and $D_j^\theta({\vec r})=e\int d\eps \df\; \nu_j^e({\vec r},\eps)$
are {\it charge injectivities} (expressed in the units of capacitances as in Ref.~[\onlinecite{Chri96}])
determined
by the energy dependent {\it particle}~\cite{Chri96} and {\it entropic}~\cite{San12} {\it injectivities}
\begin{subequations}
\begin{eqnarray}
\nu_j^p({\vec r},\eps)&=&-\frac{1}{4\pi i}\sum_k \text{Tr}\left[{\bf s}^\dagger_{kj}\frac{\delta {\bf s}_{kj}}{e\delta U({\vec r})}-\frac{\delta {\bf s}^\dagger_{kj}}{e\delta U({\vec r})}{\bf s}_{kj}\right]\label{p-inj} \, , \qquad\\
\nu_j^e({\vec r},\eps)&=&\left(\frac{\eps}{T_0}\right)\nu_j^p({\vec r},\eps) \, ,\label{e-inj}
\end{eqnarray}
\end{subequations}
where the trace runs over channel indices. Eq.~(\ref{p-inj}) gives the partial density of states at $\vec{r}$ corresponding to
injection from lead $j$, which is multiplied by a single-particle entropy factor $\eps/T_0$ in
Eq.~(\ref{e-inj})~\cite{San12}.

Next, the screening charge $\delta q^{(s)}({\vec r})$ is related to $U({\vec r})$ via the Lindhard polarization
function. The relation is in principle nonlocal, however the nonlocal corrections are quantum in essence and
thus negligible in good metals with high enough electronic density. 
We thus make the quasiclassical approximation of a local Lindhard function and write
$\delta q^{(s)}({\vec r})=-\Pi({\vec r}) U({\vec r})$, with $\Pi({\vec r}) = \sum_j D_j^{V} ({\vec r}) = D({\vec r})$.
From the definition of $D_j^V$ we have
$D({\vec r})=e^2\int d\eps \df\; \nu({\vec r},\eps)$ with the local density of states
$\nu({\vec r},\eps)=\sum_j \nu_j^p({\vec r},\eps)$. Note that at zero temperature, 
$D({\vec r})$ is the density of states (in units of capacitance) at the Fermi energy.
Putting all this together,  the net total 
injected charge is, to leading order in the biases,
\begin{equation}
\delta q({\vec r}) = \sum_j D_j^{V} ({\vec r}) V_j + \sum_j D_j^{\theta} ({\vec r}) \theta_j - D({\vec r}) U({\vec r}) \, .\label{dq_final}
\end{equation}
The conductor may also be capacitively coupled to the reservoirs and  
external metallic gates. One way to take these couplings into account is to
discretize the conductor into different regions labelled $r=1,2,...$, each with a constant 
potential $U_r$. The procedure 
brings about additional relations
\begin{equation}
\delta q_r = \sum_k C_{rk} (U_r - V_k) \, , \label{dq_capac}
\end{equation}
with the sum running over all capacitively coupled components. This equation expresses changes in local
charge distributions 
due to Coulomb interactions. The latter are represented here by geometric capacitances between
different components of the system.

Once the microstructure considered is consistently defined, in that all
electric components (leads, gates, etc.) that are coupled to the conductor are included
so that electric field lines do not leave the system, 
then the total charge is conserved by Gauss' theorem. 
Eqs.~(\ref{eq:charpot}), (\ref{dq_final})  and (\ref{dq_capac}) then determine the characteristic potentials
in Eqs.~(\ref{eq:abc}). The nonlinear transport coefficients are finally calculated
by using the identity 
$\int {\rm d} {\bf r} \, \delta_{U(\vec{r})} (...) = -e \partial_E (...)$ to 
compute the functional derivatives in Eqs.~(\ref{eq:abc}).
Two examples of thermoelectric microdevices will be discussed in 
Section~\ref{illustration}, which illustrate how the procedure is applied in practice
by discretizing the microstructures
into regions of uniform local potential. 

\section{conservations and sum rules}\label{sumrules}

Particle conservation is expressed mathematically by the unitarity of the scattering
matrix, $\sum_i A_{ij}=\sum_j A_{ij}=0$. This translates into the following
sum rules for the linear transport coefficients, 
\begin{align}
 &\sum_{i\text{ or }j} G_{ij} = \sum_{i\text{ or }j} B_{ij} = \sum_{i\text{ or }j} \Gamma_{ij} = \sum_{i\text{ or }j} \Xi_{ij} = 0 \, . \label{conservlin}
\end{align} 
In particular, two of these sum rules correspond to 
electric current conservation, $\sum_i G_{ij}=\sum_i B_{ij}=0$ while two others reflect
gauge invariance, $\sum_j G_{ij}=\sum_j \Gamma_{ij}=0$.

The unitarity of the scattering matrix further implies $\sum_i \alpha_{ijk} = \sum_i \beta_{ijk} = \sum_{i} \gamma_{ijk}=\sum_j \left(\alpha_{ijk}+\alpha_{ikj}\right) = \sum_{j} \gamma_{ijk}=0$,
which results in the following sum rules between the nonlinear response coefficients,
\begin{subequations}
\begin{align}
&\sum_i G_{ijk} = \sum_i B_{ijk} = \sum_{i} Y_{ijk} =0 \, , \label{conserv1}\\
&\sum_j \left(G_{ijk}+G_{ikj}\right) = \sum_j \left(\Gamma_{ijk}+\Gamma_{ikj}\right) =0 \, , \label{conserv2}\\
& \sum_j Y_{ijk}
= \sum_j \Psi_{ijk} =0 \, , \label{conserv3}\\
& \sum_i \Xi_{ijk} =0\, , \\
&G_{jk} = -\sum_i \Gamma_{ijk} \, , \qquad
B_{jk} = -\sum_i \Psi_{ijk} \, .\label{joule}
\end{align}
\end{subequations}
Two of these sum rules were derived in Ref.~[\onlinecite{Chri96}], $ \sum_i G_{ijk}=0$, implied by current
conservation and $\sum_j \left(G_{ijk}+G_{ikj}\right) =0$, resulting from gauge invariance. 
All other sum rules are written here for the first time. Eqs.~(\ref{conserv1})  reflect current conservation
while Eqs.~(\ref{conserv2}) and (\ref{conserv3}) correspond to gauge invariance.
The sum rules of  Eq.~(\ref{joule}) imply $\sum_i J_i + I_i^{({\rm lin})}V_i = 0$, 
where $I_i^{({\rm lin})}$ is the linear response current through lead $i$. Thus in our approach
energy is conserved
to quadratic order in the biases. 

Finally we note the following, additional double-sum rules
$\sum_{jk} \alpha_{ijk} = \sum_{jk} \beta_{ijk} = 0$ which imply
\begin{subequations}
\begin{eqnarray}
\sum_{j,k} G_{ijk} &=& \sum_{j,k} \Gamma_{ijk} =0 \, , \label{2bleimplied} \\
\sum_{j,k} B_{ijk} &=& \sum_{j,k} \Xi_{ijk} =0 \, . \label{2blenew} 
\end{eqnarray}
\end{subequations}
While Eqs.~(\ref{2bleimplied}) are guaranteed to hold by Eqs.~(\ref{conserv2}), Eqs.~(\ref{2blenew})
are new. Eqs.~(\ref{conservlin}), (\ref{conserv3}), (\ref{2bleimplied}) and (\ref{2blenew}) 
ensure in particular that the electric and heat currents of Eqs.~(\ref{eq:weaknlin}) vanish at equilibrium.

\section{Performance metrics for thermoelectricity}\label{efficiency}

For two-terminal setups treated in linear response, Eqs.~(\ref{eq:currents_initial}) reduce to 
\begin{align}\label{eq:lin1}
\begin{pmatrix}I\\J\end{pmatrix}
=
\begin{pmatrix}G&B \\ \Gamma& \Xi\end{pmatrix}
\begin{pmatrix}V\\\Delta T\end{pmatrix} \, ,
\end{align}
where $V$ and $\Delta T$ are the voltage and temperature differences between the reservoirs. Alternatively, this linear relation is often rewritten in terms  of 
responses to an electric current and a temperature difference,
\begin{align}\label{eq:lin2}
\begin{pmatrix}V\\J\end{pmatrix}
=
\begin{pmatrix}R&S \\ \pi_p& \kappa\end{pmatrix}
\begin{pmatrix}I\\\Delta T\end{pmatrix}.
\end{align}
Eqs.~(\ref{eq:lin1}) and (\ref{eq:lin2}) define the linear response coefficients, the electric 
conductance, $G$, and resistance, $R$, the Seebeck, $S$, and Peltier, $\pi_p$, coefficients
as well as the thermal conductances $\Xi$ (at zero voltage, $V=0$) and $\kappa$
(at zero current, $I=0$).
We can express the relationships between these linear transport coefficients as 
\begin{align}
\begin{pmatrix}R&S \\ \pi_p& \kappa\end{pmatrix}
=
\begin{pmatrix}1/G\quad&-B/G \\ \Gamma/G\quad& \Xi-B\Gamma/G\end{pmatrix}.
\end{align}
Dimensional analysis shows that a single dimensionless quantity can be constructed from the
linear response coefficients and the temperature. This is the figure of merit $ZT=(G S^2 / \kappa) T_0$.
In the linear regime, thermodynamic efficiencies are then only functions of $ZT$ and the temperature difference,
for instance a heat engine working in the linear regime has maximal efficiency~\cite{caveat}
\begin{align}\label{eq:zteff}
\eta_{\text{max}}&={\rm max}_V \left(\frac{IV}{J}\right)=\left(\frac{\Delta T}{T_0}\right)\frac{\sqrt{ZT + 1}-1}{\sqrt{ZT + 1}+1} \, ,
\end{align}
It is important to keep in mind that this relation is valid only in the linear
regime, and that nonlinear effects may invalidate it. Linear response may even predict 
a maximal efficiency to occur outside its range of validity (at too large a bias voltage). A description
of thermodynamic efficiency based only on $ZT$ is therefore inappropriate in the nonlinear regime. 
However, the efficiency is generically given by
$\eta  =  IV/J_{\rm hot}$, 
where $J_{\rm hot}$ is the heat current flowing from the hot reservoir. While $J_{\rm hot}=-J_{\rm cold}$
in linear response, $J_{\rm hot} \ne -J_{\rm cold}$ in the nonlinear regime. Note that 
with this definition, it makes sense to substitute $T_0 \rightarrow T_{\rm hot}$ in Eq.~(\ref{eq:zteff}),
in which case one obtains Carnot efficiency for $ZT \rightarrow \infty$.

The efficiency of thermoelectric refrigeration is measured by the coefficient of performance (COP)
which is the ratio of the heat extracted from the cold reservoir to the work done~\cite{Mah97}, 
$\COP=J_{\text{cold}}/IV$. This refrigeration ``efficiency" is limited by $\COP<T_{\rm cold}/\Delta T$.
In the linear regime, its maximum is also given by $ZT$ as
\begin{equation}
\COP_{\text{max}}={\rm max}_V \left( \frac{J}{IV} \right)=
\left(\frac{T_0}{\Delta T}\right)\frac{\sqrt{ZT + 1}-1}{\sqrt{ZT + 1}+1} \, ,
\end{equation}
where this time it makes sense to substitute $T_0 \rightarrow T_{\rm cold}$.
Note that the linear response scattering theory leaves the choice of $T_0$ arbitrary, up to the
fact that a well chosen $T_0$ (close to both $T_{\rm cold}$ and $T_{\rm hot}$) minimizes
nonlinear corrections. 

It is not at all
guaranteed that $ZT$ remains a valid figure of merit once nonlinear effects kick in.
Below we discuss two examples of 
mesoscopic thermoelectric engines and compare their true efficiency to that predicted
by linear response. 

\section{examples}\label{illustration} 

\subsection{Resonant Tunneling Barrier}

We illustrate our theory, first by calculating the thermodynamic efficiency of thermoelectric
devices based on a coherent single-level double-barrier system - a resonant tunneling barrier
(RTB).
Nonlinear electric transport through a RTB was investigated 
in Ref.~[\onlinecite{Chri96}], where rectification effects were stressed in asymmetric systems with 
different tunneling probabilities through the left and right barriers. Electric transport generated by 
simultaneously applied voltage and temperature biases was investigated in Ref.~[\onlinecite{San12}],
focusing on the thermopower. This system is especially interesting in that the weakly nonlinear
theory can be compared to fully nonlinear solutions~\cite{Chri96}.

\begin{figure}[ht]
\psfrag{Ef}{$E_F$}
\psfrag{Er}{$E_r$}
\psfrag{g1}[][][1.5]{$\GL$}
\psfrag{g2}[][][1.5]{$\GR$}
\psfrag{U}{$eU$}
\psfrag{Vg}[][][1.5]{$V_g$}
\psfrag{m1}{$\mu_{\rm L}$}
\psfrag{m2}{$\mu_{\rm R}$}
\psfrag{T1}{$T_{\rm L}$}
\psfrag{T2}{$T_{\rm R}$}
\psfrag{C}[][][1.2]{$C$}
\includegraphics[width=2in]{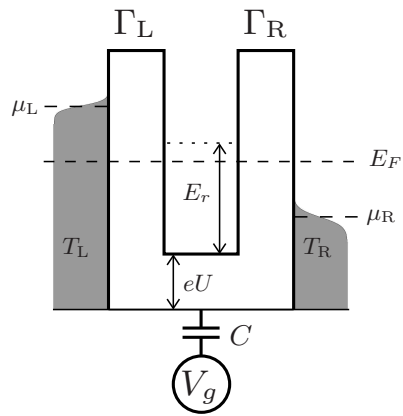}
\caption{Sketch of a resonant tunneling barrier (RTB) formed between two Fermi liquid reservoirs
by two tunnel barriers in series with transparencies $\Gamma_{\rm L}$ and $\Gamma_{\rm R}$. 
They define a
potential well with a single resonant level capacitively coupled to an external  gate. 
}\label{fig:rtb}
\end{figure}

The system we consider is sketched in Fig.~\ref{fig:rtb}. Two tunnel barriers with tunnel rates
$\Gamma_{\rm L,R}$ define a potential well with a single resonant level, at energy $E_r$ above the bottom
of the well. Left and right, the well is coupled to two Fermi liquid
reservoirs at temperatures $T_{\rm L,R}$ and electrochemical potentials $\mu_{\rm L,R}$.
The well is capacitively coupled to an external gate with voltage $V_g$ which, together with the
temperatures and electrochemical potentials in the reservoirs determine
the potential $U$ (assumed spatially homogeneous) in the well. 
This models a quantum dot in the Coulomb blockade regime as long as neither the temperatures nor the
voltage biases exceed the single-particle level spacing in the dot.
The equilibrium energy dependent particle injectivity at lead $j=$L, R is~\cite{Chri96}
\begin{align} \label{eq:injectRTB}
\nu_j^p(\eps) = \frac{1}{\pi} \frac{\Gamma_j}{(\eps+E_F-E_r)^2+\Gamma^2}\, ,
\end{align}
and the transmission probability is~\cite{Chri96}
\begin{align}
\mathcal{T}(\eps) = \frac{4 \GL \GR}{(\eps+E_F-E_r)^2+\Gamma^2} \, ,
\end{align}
where $\Gamma=\GL+\GR$ is the total decay rate of the resonant level and we defined
$E_F=(\mu_{\rm L}+\mu_{\rm R})/2$. 
In the following we will set $E_F=0$, as the zero of the energy scale
and measure all other energies  in units of $\Gamma={\rm const}$.

\begin{figure}[b] 
\includegraphics[width=8cm]{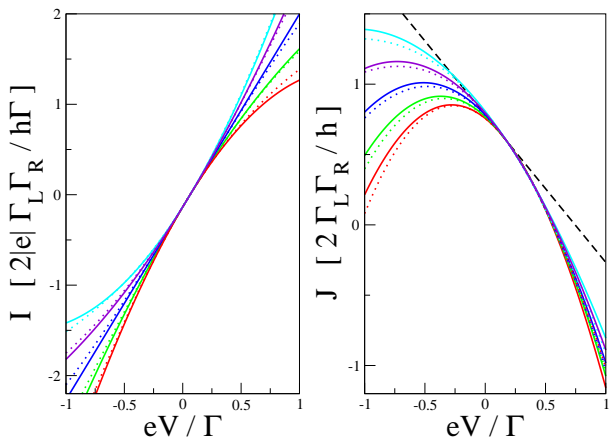} 
\caption{Electric (left) and heat (right) currents from a hot reservoir to a
RTB with $(E_r-E_F)/\Gamma=1$, $\mu_{\rm L}-\mu_{\rm R}=eV$ and $\TR-\TL=0.3\;T_0$. The base 
temperatures are
$k_B T_0/\Gamma=0.3$ (left) and $k_B T_0/\Gamma=1$ (right). The different curves correspond to
different tunnel barrier asymmetries $(\GL-\GR)/\Gamma = 0.95$ (red), 0.5 (green), 0 (blue), -0.5 (violet), -0.95 (cyan). Here, we switch off the capacitive coupling between the RTB and the external gate, $C=0$.
The solid lines are the approximate solutions in our weakly nonlinear approach and the dotted lines are the fully nonlinear solutions. The black dashed line is the heat current predicted from linear response.}\label{fig:rtb_IJ}
\end{figure}
Because we consider a single homogeneous potential in the well, $U(\vec{r}) = U$,
the injected charge loses its spatial dependence, $\delta q(\vec{r}) \rightarrow \delta q$. Consequently,
Eq.~(\ref{dq_final}) can be rewritten as
\begin{align} \label{lin_dq}
\delta q &= \sum_j D_j^{V} V_j + \sum_j D_j^{\theta} \theta_j - D U \, . \end{align}
The well is capacitively coupled to an external gate, thus
Eq.~(\ref{dq_capac}) 
tells us that the additional charge in the potential well is also given by
\begin{align}
\delta q = C(U - V_g) \, . \label{cap}
\end{align}
Using Eqs.~(\ref{lin_dq}) and (\ref{cap}) we obtain the characteristic potentials as
\begin{align} \label{char_potent}
u_j=\frac{D_j^V}{C+D},\;\; u_g=\frac{C}{C+D},\;\; z_j=\frac{D_j^\theta}{C+D} \, , 
\end{align}
where the charge and entropic injectivities are given by
$D_j^V=e^2\int d\eps \df\; \nu_j^p(\eps) $ and $D_j^\theta=e\int d\eps \df\; 
(\eps/T_0) \, \nu_j^p(\eps)$, with the particle injectivity $\nu_j^p$ of
Eq.~(\ref{eq:injectRTB}). This fully determines the nonlinear coefficients in Eqs.~(\ref{eq:nonlincoeff}) so that 
currents can be calculated using Eqs.~(\ref{eq:weaknlin}).

\begin{figure}[t]
\includegraphics[width=8.5cm]{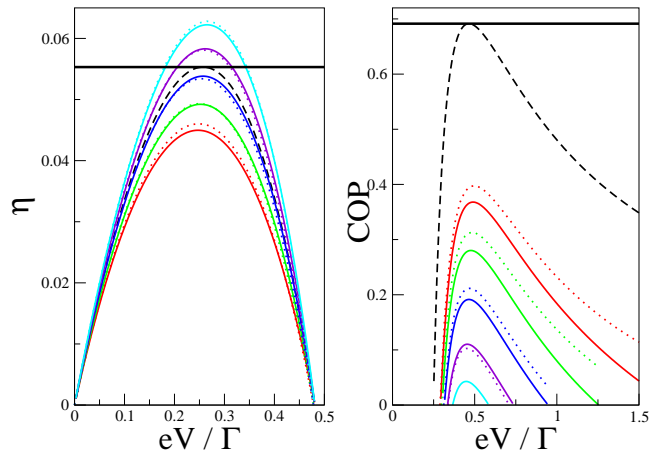}
\caption{Left panel: Thermodynamic efficiency for a RTB-based heat engine with
$(E_r-E_F)/\Gamma=1$,  $\mu_{\rm L}-\mu_{\rm R}=eV$, $k_B T_0/\Gamma=1$, $\TR-\TL=0.8\;T_0$ and 
different 
asymmetries $(\GL-\GR)/\Gamma = 0.95$ (red), 0.5 (green), 0 (blue), -0.5 (violet), -0.95 (cyan). 
Right panel: COP for the same RTB as in the left panel with $\TR-\TL=0.1\;T_0$ and
working at larger bias, thus as a refrigerator. 
In both cases
there is no capacitive coupling between the RTB and the external gate, $C=0$.
The solid lines are the weakly nonlinear approximate solutions and the dotted lines are the fully nonlinear solutions (see text).
The dashed black line is the linear response solution and the solid black line is the maximum efficiency/COP predicted from linear response.}\label{fig:rtb_eff_sym}
\end{figure}
For the RTB the potential $U$ can alternatively be determined nonperturbatively.
Eq.~(\ref{lin_dq}) for
the injected charge is replaced by the energy integral
\begin{align} \label{eq:exactrtb}
\delta q &=e\int d\eps \;\left[\sum_j \; \nu_j^p(\eps-eU) \; f_j(\eps) - \nu(\eps) \; f(\eps)\right]\, ,
\end{align}
with $\nu(\eps) = \sum_j \nu_j^p(\eps)$. 
One determines $U$ numerically
from Eqs.~(\ref{cap}) and (\ref{eq:exactrtb}).
The fully nonlinear currents are then obtained from 
Eqs.~(\ref{eq:currents_initial}). The result is exact, up to the approximation of a single homogeneous
local potential in the well. 

We first  show in Fig.~\ref{fig:rtb_IJ} the electric and heat currents for two RTB's. Different temperatures
have been chosen to present $I$ and $J$ in order to visually optimize nonlinear effects.
The curves are clearly nonlinear and display rectification in both the
electric (left) and heat (right) currents. The rectification in the heat current is in general 
more pronounced, due to the 
additional voltage in the integrand of Eq.~(\ref{eq:1b}). Most remarkably, perhaps, we see that the 
behavior obtained via the weakly nonlinear approximation (solid lines) 
does not differ much from the fully nonlinear result (dotted lines), even for the strongly nonlinear 
heat current. The pronounced rectification effects will turn out to boost or hinder the efficiency of RTB-based 
devices depending on the asymmetry $(\GL-\GR)/\Gamma$.

We next investigate the efficiency of RTB-based thermoelectric devices. 
Fig.~\ref{fig:rtb_eff_sym} shows 
the thermodynamic efficiency, $\eta=IV/J_{\rm hot}$, of a heat engine (left panel) and 
the coefficicent of performance, $\COP=J_{\rm cold}/IV$, of a refrigerator in the linear (dashed black line)
and in the nonlinear approximation (colored solid and dotted lines) for different
asymmetries $(\GL-\GR)/\Gamma$. In linear response, $I \propto \GL \GR/\Gamma$
and $J \propto \GL \GR$, so that both $\eta$ and $\COP$ are independent of 
$(\GL-\GR)/\Gamma$. For the chosen parameters,
the maximal efficiency is at finite voltage bias, yet linear response gives a  good approximation to
$\eta$ in the symmetric case, $\GL=\GR$. This is so, because 
in that case, $D_{\rm L}^{V,\theta}=D_{\rm R}^{V,\theta}$ and the screening potential given by 
Eq.~(\ref{lin_dq}) identically vanishes in the absence of gate coupling, $C=0$.
Introducing an asymmetry in the RTB changes the rules of the
game, as rectification in the electric and heat currents kick in. 
As the asymmetry increases we can see a clear deviation from linear response, which emphasizes 
the importance of treating screening effects when characterizing thermoelectric performances of such
microdevices.
The efficiency $\eta$ increases or decreases, 
depending on whether rectification play with or against the bias voltage.
This depends on the sign of the bias voltage and the direction of rectification.
From Fig.~\ref{fig:rtb_IJ} we see that at positive voltage,
samples with  large negative values for the asymmetry $(\GL-\GR)/\Gamma$ exhibit 
rectification effects where simultaneously  the electric current is enhanced while  the heat current
is reduced
compared to their values in the linear regime. Both effects conspire to increase the efficiency
in that case (cyan curves in the left panel of Fig.~\ref{fig:rtb_eff_sym}). 
The coefficient of performance, on the other hand, is always significantly smaller than predicted
by linear response, and rectification plays in the opposite direction in
the refrigerator case because $\COP = J_{\rm cold}/IV$ while $\eta=IV/J_{\rm hot}$.
Another remarkable
feature is that both performance metrics are well predicted by the weakly nonlinear theory (compare
solid with dotted lines) with larger deviations, as expected, in the refrigerator case at larger
biases.

\begin{figure}[t]
\includegraphics[width=8.5cm]{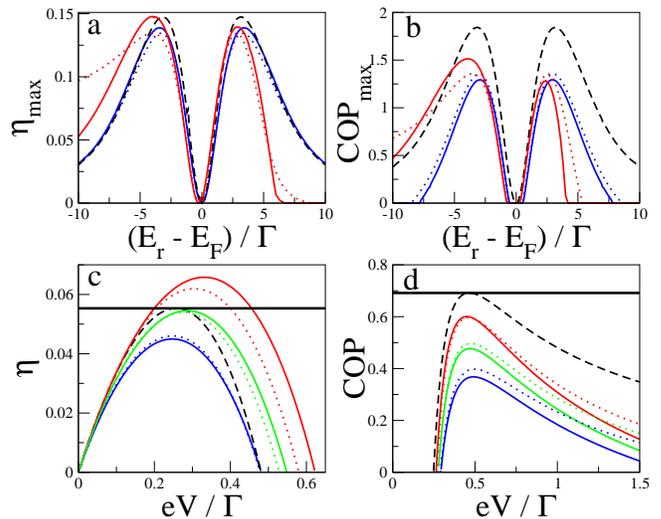}
\caption{a) Maximal thermodynamic efficiency, $\eta_{\rm max} = {\rm max}_V \eta$,
for a RTB heat engine and b)
maximal coefficient of performance, $\COP_{\rm max} = {\rm max}_V \COP$, for
a RTB refrigerator as a function of the
position of the resonant level.  
c) Efficiency and d) coefficient of performance vs bias voltage
for RTB-based devices with $(E_r-E_F)/\Gamma=1$.
In all cases, $k_B T_0/\Gamma=1$, $(\GL-\GR)/\Gamma = 0.95$ and
the coupling to the gate voltage varies as
$C \Gamma/e^2=0$ (blue), $5\cdot 10^{-6}$ (green), and $10^{-5}$ (red) with a gate voltage given by $eV_g/\Gamma =3000$. Panels a) and c) are for $\TR-\TL=0.8\;T_0$ and panels b) and d) are for $\TR-\TL=0.1\;T_0$.
The solid colored lines are the approximate solutions in our weakly nonlinear approach and the dotted 
colored lines are the fully nonlinear numerical solutions. The dashed black line is the linear response solution 
and the solid black line is the maximum efficiency/COP predicted from linear response. }\label{fig:rtb_eff}
\end{figure}

So far the capacitive coupling to the external gate has been set equal to zero, $C=0$.
Fig.~\ref{fig:rtb_eff} finally shows the dependence of $\eta$ and $\COP$ on  $C$. 
Efficiencies and maximal efficiencies for a heat engine are shown in the left
panels, while coefficients of performance and maximal coefficients of performance
for a refrigerator are on the right panels. For the chosen set
of parameters, linear response overestimates $\eta$ and $\COP$ 
when there is no coupling to the gate.
In both panels c) and d), the capacitive
coupling is seen to increase $\eta$ and $\COP$. Quite remarkably
the efficiency ranges from well below the linear prediction for $C \Gamma/e^2=0$ to well above it
for $C \Gamma/e^2=10^{-5}$, strikingly illustrating the dependence of $\eta$ on nonlinear effects.
The same calculation for RTB's with all other asymmetries reported in Fig.~\ref{fig:rtb_eff_sym}
show systematic increases in $\eta$  above its value at $C \Gamma/e^2=0$ as the capacitive coupling
is turned on. The magnitude of the increase is in all cases similar to that in Fig.~\ref{fig:rtb_eff},
which we attribute to similar increases in $U$ with $C$, regardless of the asymmetry.
Panels a) and b) also show that better performances are obtained for a resonant level
lying a distance of a few times $\Gamma$ above or below $E_F$.
We finally note that,
as before and despite the 
rather large biases involved, the weakly nonlinear theory gives a rather faithful prediction of the
true fully nonlinear behavior.

From the data shown in this section we conclude that nonlinear effects can significantly increase
the efficiency of RTB-based thermoelectric engines, first via rectification effects in both the
electric and heat currents, and second via capacitive couplings which lift the resonant level.

\subsection{Quantum Point Contact}

\begin{figure}[t]
\includegraphics[width=8cm]{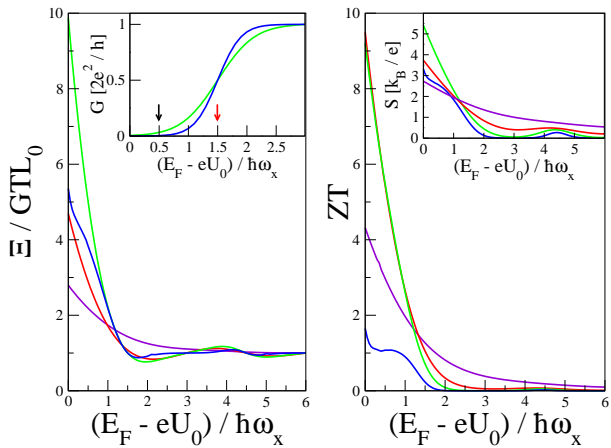}
\caption{Left panel: dimensionless ratio $\Xi/G T L_0$ for a QPC as defined in the text. 
This ratio is equal to one when the 
Wiedemann-Franz law is valid. Right panel: 
dimensionless figure of merit $ZT$. Left inset: conductance steps for the QPC. The arrows indicate the
parameters corresponding to Figs.~\ref{fig:qpc_eff} and \ref{fig:qpc_eff2}.
Right inset: thermopower $S$ of a QPC. Parameters are 
$\omega_y/\omega_x=3$ and $T=0.10$ (blue), 0.25 (green), 0.50 (red), 1.00 (violet).}\label{fig:linqpc}
\end{figure}

Nonlinear electric transport through a quantum point contact (QPC) has been discussed
theoretically in Ref.~[\onlinecite{Chri96}] and experimentally in Ref.~[\onlinecite{Pat91}]. 
Linear thermoelectric transport through a QPC has been discussed by Proetto~\cite{Pro91},
assuming a saddle potential whose transmission spectrum is known exactly~\cite{Butt90,Con68,Mil68}.
Oscillations in the thermopower were in particular reported in Ref.~[\onlinecite{Pro91}],
which we reproduce in the inset to the right
panel of Fig.~\ref{fig:linqpc}. Curiously enough, the efficiency of QPC-based thermoelectric
devices has not been calculated as of yet, even in the linear regime, and we briefly comment
first on the figure of merit for a QPC.
We show in Fig.~\ref{fig:linqpc} data for the 
ratio $\Xi/G T L_0$ with the Lorenz number $L_0=\pi^2 (k_{\rm B}/e)^2/3$
as well as $ZT = (G S^2 / \kappa) T_0$. When the Wiedemann-Franz law holds, one has
$\Xi/G T L_0=1$~\cite{Ashcroft}. 
This is the case when the QPC is well opened and transmits several 
conductance channels.  
The Wiedemann-Franz law breaks down when the QPC is closing, and the 
sharp increase displayed by $\Xi/G T L_0$ in that regime suggests a reduction in
$ZT$. The right panel of Fig.~\ref{fig:linqpc} however indicates an equally sharp
increase of $ZT$ reaching almost 10, which is understood once one realizes (see inset) 
that the increase in $\Xi/G T L_0$ is accompanied by an increase of $S$~\cite{Pro91}, 
with the net result
that $ZT$ increases as the QPC closes. This results in a large efficiency for an almost
completely closed QPC operating as a heat engine. This  however occurs when
the power of the device is exponentially small. Some aspects of nonlinear thermoelectric transport 
through a QPC have been addressed in Ref.~[\onlinecite{Rob12.1}]. We here extend these investigations.

\begin{figure}[t]
\psfrag{V}[][][1.5]{$V_g$}
\psfrag{u1}[][][1.5]{$\UL$}
\psfrag{u2}[][][1.5]{$\UR$}
\psfrag{C}{$C$}
\psfrag{Co}{$C_0$}
\includegraphics[width=5.5cm]{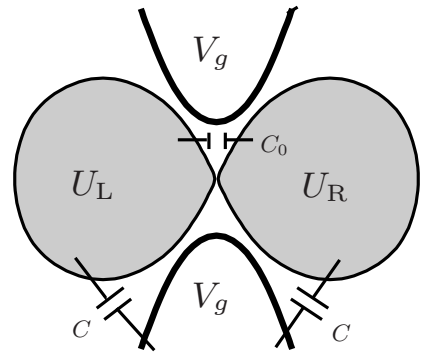}
\caption{Sketch of a quantum point contact (QPC) 
with the local potential in the conductor discretized into a left and
a right region. The QPC is electrostatically defined by external gates. The capacitive couplings considered
in our theoretical treatment are also indicated.}\label{fig:qpc}
\end{figure}

The geometry of the QPC is sketched in Fig.~\ref{fig:qpc}. Two lateral gates constrain the motion of 
electrons in a two-dimensional electron gas to a narrow channel. We take these two gates as symmetric
and at the same potential and thus consider them together as a single gate.
The width of the channel is tuned
by the electric potential on the gates. We spatially split the QPC into a left and a right region,
each with its own local potential. These two regions are capacitively
coupled to one another (capacitance $C_0$) and to the gate. For physically realistic parameters,
an asymmetry in the capacitive coupling of the two regions to the gate has only a small effect on transport,
and we therefore consider a symmetric coupling with capacitance $C$ for both the left and right
regions of the QPC. We consider a
geometrically  symmetric QPC with only a few open transmission channels, which we model 
by a saddle potential~\cite{Butt90} 
\begin{align}
eU(x,y) &= eU_0 + \frac{1}{2}m \omega_y^2 y^2 - \frac{1}{2}m \omega_x^2 x^2 \, ,
\end{align}
with the electron mass $m$ and the electric potential $U_0$ at the center of the QPC.
The transmission probability for each transmission channel is given by~\cite{Con68,Mil68}
\begin{align}
\mathcal{T}_n(\eps) &= \left[1+e^{-\pi \epsilon_n (\eps)}\right]^{-1}  \, , 
\end{align}
with
\begin{align}
\epsilon_n (\eps) &= 2\left[\eps+E_F - \hbar\omega_y\left(n+\frac{1}{2}\right)-eU_0\right]\Big/\hbar \omega_x,
\end{align}
for $n=0,1,2,\ldots$
The density of states for each channel, $\nu_n(\eps)$, 
has been calculated in Ref.~[\onlinecite{Ped97}], using a WKB approach. The result exhibits a
nonphysical logarithmic singularity close to conductance steps, but 
retains its validity away from them. We found that electric and heat currents
lie on smooth curves except in the immediate vicinity of the steps. The results presented below
accordingly use a numerical interpolation to investigate the behavior of QPC-based devices close
to conductance steps,
whose exact form does not matter, given the very small range where 
the expression of Ref.~[\onlinecite{Ped97}] breaks down. Other than this interpolation, 
we use the density of states calculated in Ref.~[\onlinecite{Ped97}].

The QPC has been discretized into two separate regions, and we express the particle injectivity into region $r={\rm L},{\rm R}$ from lead $j=$L, R as~\cite{Chri96} 
\begin{align}
\nu^p_{rj}(\eps) = \frac{1}{2} \sum_n \nu_n(\eps) \,  \left[\mathcal{T}_n(\eps) + 2 [1-\mathcal{T}_n(\eps)] \delta_{rj}\right] \, ,
\end{align}
where the prefactor of $1/2$ 
takes into account that only
electrons moving toward the QPC undergo transmission or reflection.
Eq.~(\ref{dq_final}) now becomes
\begin{align} 
\delta q_r &\simeq \sum_j D_{rj}^{V} V_j + \sum_j D_{rj}^{\theta} \theta_j - D U_r \, , \label{qpcDQ1}
\end{align}
where $r={\rm L,R}$ labels the left/right regions of the QPC. As in Eq.~(\ref{cap}), the injected charge can
alternatively be determined by 
\begin{align}
\delta q_{r} = C(U_{r} - V_g) + C_0 (U_{r}-U_{\bar r}), \label{qpcDQ2}
\end{align}
where ${\bar r}=L(R)$ when $r=R(L)$.
Eqs.~(\ref{qpcDQ1}) and (\ref{qpcDQ2}) determine the characteristic potentials as
\begin{alignat}{2}	
u_{rj} &= \frac{(C+C_0+D)D_{rj}^V+C_0 D_{{\bar r}j}^V}{(C+D)(C+2C_0+D)}, & \;\; u_{rg} &= \frac{C}{C+D}\\
z_{rj} &= \frac{(C+C_0+D)D_{rj}^\theta+C_0 D_{{\bar r}j}^\theta}{(C+D)(C+2C_0+D)}, &z_{rg} &= 0\, ,
\end{alignat}
which, again, we use to calculate nonlinear transport coefficients in Eqs.~(\ref{eq:nonlincoeff}) and 
the currents in Eqs.~(\ref{eq:weaknlin}). Note that the characteristic potentials now have two labels, one indicating the
regions where they are measured, the other one the lead or gate from which they originate.

\begin{figure}[t]
\includegraphics[width=8.5cm]{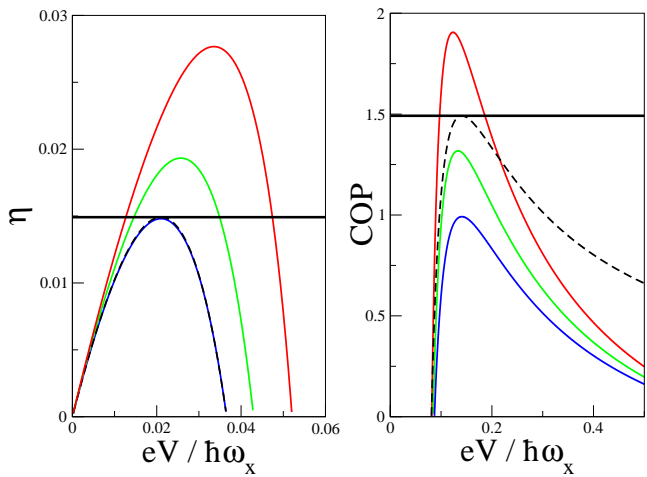}
\caption{Left panel: Efficiency of a QPC-based heat engine. Right panel: Coefficient of 
performance for a QPC-based refrigerator. In both instances the QPC is set at
 $(E_F-eU_0)/\hbar \omega_x=1.5$ corresponding to the red arrow in the inset of 
 Fig.~\ref{fig:linqpc},
 $\omega_y/\omega_x=3$, $eV_g/\hbar\omega_x=3000$,  $k_B T_0/\hbar \omega_x = 0.3$, 
 $\TR-\TL=0.1\;T_0$, $eV=E_F-\mu_{\rm R}=\mu_{\rm L}-\mu_{\rm R}$, 
 $C_0=0$ and $C \hbar\omega_x/e^2=0$ (blue), $5\cdot 10^{-5}$ (green), and $10^{-4}$ (red). }\label{fig:qpc_eff}
\end{figure}

We present our results for the efficiency of QPC-based devices. For all presented data, we set the 
chemical potential of the left terminal at $E_F$ and bias the right reservoir to lower electrochemical
potential, $\mu_{\rm R} = E_F-eV$. This corresponds approximately to the situation considered
by Whitney~\cite{Rob12.1} (see his Fig.2d), though he chose a Heaviside function for 
the QPC transmission. Another discrepancy between that work and ours is that we discretize the geometry into
two distinct spatial regions, which accordingly may carry two different local potentials. 
The data we present all have $C_0=0$, as we found that
this capacitive coupling does not change the results significantly. Note that for the QPC, only the
weakly nonlinear results are shown.

In Fig.~\ref{fig:qpc_eff} we plot both the efficiency of a QPC-based heat engine (left panel) and
the coefficient of performance of a QPC-based refrigerator (right panel) 
 as a function of the applied bias voltage when the capacitive coupling $C$ to the gate
 is varied. Both sets of data correspond to a  first transmission
 channel close to half transmission, $\mathcal{T} \simeq 1/2$, corresponding to the red arrow in the
 inset of the left panel of Fig.~\ref{fig:linqpc}. 
Fig.~\ref{fig:qpc_eff} shows
that in the absence of capacitive coupling,
the maximal efficiency of the heat engine is slightly lower than one would predict from linear response, 
in agreement with Ref.~[\onlinecite{Rob12.1}]. The trend is however reversed once the capacitive coupling  $C$
is turned on. Increasing $C$ increases $\eta$ and even extends the bias range 
of  operation of the heat engine towards higher bias. Turning our attention next to the refrigerator case
(right panel in Fig.~\ref{fig:qpc_eff}) we note that the
discrepancy between linear prediction and nonlinear calculation is stronger, essentially because
the refrigerator  works at higher bias voltage, therefore larger nonlinear effects may be expected.
Increasing $C$ increases the COP for refrigeration just as it increased the power generation efficiency for the QPC when operating as a heat engine. In both instances we see that the nonlinear efficiency 
significantly exceeds the linear prediction at large enough capacitive coupling $C$. 

\begin{figure}[t]
\includegraphics[width=8.5cm]{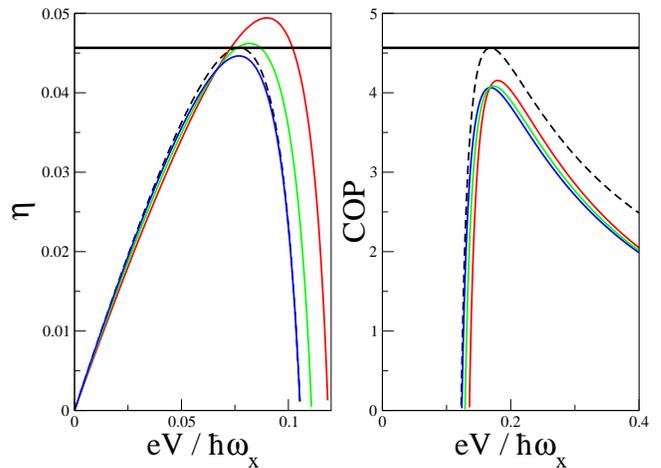}
\caption{Left panel: Efficiency of a QPC-based heat engine. Right panel: Coefficient of 
performance for a QPC-based refrigerator. In both instances the QPC is set at
 $(E_F-eU_0)/\hbar \omega_x=0.5$ corresponding to the black arrow in the inset of 
 Fig.~\ref{fig:linqpc},
 $\omega_y/\omega_x=3$, $eV_g/\hbar\omega_x=3000$,  $k_B T_0/\hbar \omega_x = 0.3$, 
 $\TR-\TL=0.1\;T_0$, $eV=E_F-\mu_{\rm R}=\mu_{\rm L}-\mu_{\rm R}$, 
 $C_0=0$ and $C \hbar\omega_x/e^2=0$ (blue), $2\cdot 10^{-5}$ (green), and $4\cdot 10^{-5}$ (red). }\label{fig:qpc_eff2}
\end{figure}

Fig.~\ref{fig:linqpc} shows a dramatic increase of $ZT$ when the QPC is almost totally closed and its
conductance is exponentially small. We investigate whether this region presents opportunities
for thermoelectricity, in particular whether the already large linear efficiency can be further increased
by nonlinear effects. In Fig.~\ref{fig:qpc_eff2} we plot data similar to those presented in 
Fig.~\ref{fig:qpc_eff} but obtained at smaller $E_F$, corresponding to the black arrow in the
 inset of the left panel of Fig.~\ref{fig:linqpc}. We see that nonlinearities influence thermodynamic
 performances only weakly in that regime, essentially because currents are exponentially small.

\section{Conclusions}\label{conclusions}

We have presented a weakly nonlinear theory of coupled heat and electric transport in 
micro- and nanoscopic devices. On those scales it is of central importance that local
electrostatic potentials are correctly taken care of, in order to preserve fundamental 
conservation laws such as gauge invariance and current conservation. We described how this
can be done and listed sum rules ensuring that these conservation laws are preserved. 
Most works on thermoelectricity use the linear figure of merit $ZT$
as a metric for thermodynamic performance -- the larger $ZT$, the better the efficiency of a heat
engine and the coefficient of performance of a refrigerator. We pointed out that the predictive
power of $ZT$ is no longer guaranteed when nonlinear effects kick in, and directly investigated
the thermodynamic efficiency and coefficient of performance, including nonlinear
contributions to the heat and electric currents. We investigated devices based on two
microconstrictions, a quantum point contact and a resonant tunneling barrier. In both instances
we found situations where the true efficiency significantly differs from that predicted by linear
response, as well as cases where thermodynamic efficiencies strongly depend on 
the strength of these nonlinearities -- not only on the magnitude of the applied temperature and
voltage biases, but also on the capacitive coupling to external electrostatically charge regions. 
Most importantly, we found that the weakly nonlinear theory faithfully captures the true
nonlinear behavior of micro- and nanoscopic devices. We believe this opens the way to
more general investigations of thermoelectricity in quantum microdevices which until now
have often been limited to numerical investigations.

{\it Acknowledgments.} 
This work was supported by the NSF under grant 
PHY-1001017. We thank Andrew Jordan for drawing our
attention to Ref.~[\onlinecite{San12}].
While this work was being completed, we received a draft of 
Ref.~[\onlinecite{Rob12.2}] which discusses related issues in nonlinear 
transport.


\begin{thebibliography}{99}
\bi{Chu12} S. Chu and A. Majumdar, Nature {\bf 488}, 294 (2012).

\bi{Mah97} G. Mahan, B. Sales, and J. Sharp, Phys. Today {\bf 50}, vol. 3,
42 (1997).

\bi{Maj04} A. Majumdar, Science {\bf 303}, 777 (2004).

\bi{Sny08} G.J. Snyder and E.S. Toberer, Nat. Materials {\bf 7}, 105 (2008).

\bi{Sha11} A. Shakouri, Ann. Rev. Mat. Res. {\bf 41}, 399 (2011).

\bi{Dres07} M.S. Dresselhaus,
G. Chen, M.Y. Tang, R. Yang, H. Lee, D. Wang, Z. Ren, J.-P. Fleurial,
and P. Gogna, Adv. Mater. {\bf 19}, 1043 (2007).

\bi{Mah96} G.D. Mahan and J.O. Sofo, PNAS {\bf 93}, 7436 (1996).

\bi{Gia06} F. Giazotto, T.T. Heikkil\"a, A. Luukanen, 
A.M. Savin, and J.P. Pekola, Rev. Mod. Phys. {\bf 78}, 217 (2006).

\bi{Sch08} R. Scheibner, M. K\"onig, D. Reuter, A.D.Wieck, C. Gould,
H. Buhmann, and L.W. Molenkamp, 
New J. Phys. {\bf 10}, 083016 (2008).

\bi{Sot12} B. Sothmann, R. Sanchez, A.N. Jordan, and M. B\"uttiker,
Phys. Rev. B {\bf 85}, 205301 (2012).

\bi{Mat12} J. Matthews, D. Sanchez, M. Larsson, and H. Linke, 
Phys. Rev. B {\bf 85}, 205309 (2012).

\bi{Edw95} H.L. Edwards, Q. Niu, G.A. Georgakis, and A.L. de Lozanne,
Phys. Rev. B {\bf 52}, 5714 (1995).

\bi{Pra09} J.R. Prance, C.G. Smith, J.P. Griffiths, S.J. Chorley, D.
Anderson, G.A.C. Jones, I. Farrer, and D.A. Ritchie, 
Phys. Rev. Lett. {\bf 102}, 146602 (2009).

\bibitem{Rob12.1}  R.S. Whitney, arXiv:1208.6130.

\bi{Hum02} T.E. Humphrey, R. Newbury, R.P. Taylor, and
H. Linke, Phys. Rev. Lett. {\bf 89}, 116801 (2002).

\bi{San11} R. Sanchez and M. B\"uttiker, 
Phys. Rev. B {\bf 83}, 085428 (2011).

\bi{Mur12} B. Muralidharan and M. Grifoni, 
Phys. Rev. B {\bf 85}, 155423
(2012).

\bi{Red07} P. Reddy, S.Y. Jang, R.A. Segalman, and A. Majumdar,
Science {\bf 315}, 1568 (2007).

\bi{Bah08} K. Baheti, J.A. Malen, P. Doak, P. Reddy, S.Y. Jang,
T.D. Tilley, A. Majumdar, and R.A. Segalman, Nano Lett. {\bf 8}, 715 (2008).

\bi{Berg09} J.P. Bergfield and C.A. Stafford,  Nano Lett. {\bf 9}, 3072 (2009).

\bi{Noz10}  D. Nozaki, H. Sevincli, W. Li, R. Guti\'errez, and G. Cuniberti,
 Phys. Rev. B {\bf 81}, 235406 (2010).

\bi{Ent10} O. Entin-Wohlman, Y. Imry, and A. Aharony, Phys. Rev. B {\bf 82}, 115314 (2010).

\bi{Yee11} S.K. Yee, J.A. Malen, A. Majumdar, and R.A. Segalman, 
Nano Lett. {\bf 11}, 4089 (2011).

\bibitem{Nik12} B.K. Nikoli\'c, K.K. Saha, T. Markussen, and K.S. Thygesen, 
J. Comput. Electron. {\bf 11}, 78 (2012);
K.K. Saha, T. Markussen, K.S. Thygesen, and B.K. Nikoli\'c,
Phys. Rev. B {\bf 84}, 041412(R) (2011).


\bi{Ashcroft} N.W. Ashcroft and N.D. Mermin, {\it Solid State Physics}, Saunders (Philadelphia, 1976).

\bi{hill01} R.W. Hill, C. Proust, L. Taillefer, P. Fournier, and R.L. Greene,
Nature {\bf 414}, 711 (2001).

\bi{Col05} P. Coleman, J.B. Marston, and A.J. Schofield,
Phys. Rev B {\bf 72}, 245111 (2005).

\bi{Pod07} D. Podolsky, A. Vishwanath, J. Moore, and S. Sachdev,
Phys. Rev B {\bf 75}, 014520 (2007).

\bi{Vav05} M.G. Vavilov and A.D. Stone, Phys. Rev. B {\bf 72}, 205107 (2005).

\bi{Bal12} V. Balachandran, R. Bosisio, and G. Benenti, Phys. Rev. B {\bf 86}, 035433 (2012).

\bi{Jian05} Z. Jiang and V. Chandrasekhar, 
Phys. Rev. B {\bf 72}, 020502(R) (2005).

\bi{Vir07} P. Virtanen and T.T. Heikkil\"a, Appl. Phys. A {\bf 89}, 625 (2007).

\bi{RobTP} Ph. Jacquod and R.S. Whitney,
Europhys. Lett. {\bf 91}, 67009 (2010).

\bi{Dzu93} A.S. Dzurak, C.G. Smith, L. Martin-Moreno, M. Pepper, D.A. Ritchie,
G.A.C. Jones, and D.G. Hasko, J. Phys.: Condens. Matter {\bf 5}, 8055 (1993).



\bi{Sho01} I. Shorubalko, H. Q. Xu, I. Maximov, P. Omling, L. Samuelson, and W. Seifert, Appl. Phys. Lett. {\bf 79}, 1384 (2001).

\bi{Cas02} M. Terraneo, M. Peyrard, and G. Casati, Phys. Rev. Lett. {\bf 88},
094302 (2002).

\bi{San04} D. Sanchez and M. B\"uttiker, Phys. Rev. Lett. {\bf 93}, 
106802 (2004).

\bi{Spi04} B. Spivak and A. Zyuzin, 
Phys. Rev. Lett. {\bf 93}, 226801 (2004).

\bibitem{And06} A.V. Andreev and L.I. Glazman, Phys. Rev. Lett. {\bf 97}, 266806 (2006).

\bi{Mea12} J. Meair and Ph. Jacquod, J. Phys.: Condens. Matter {\bf 24}, 272201 (2012).


\bi{Gri91} A.N. Grigorenko, P.I. Nikitin, D.A. Jelski, and T.F. George, 
Phys. Rev. B {\bf 42}, 7405  (1990).

\bi{Kul94} I.O. Kulik, J. Phys. condens. Matter {\bf 6}, 9737 (1994).

\bi{Free06} J.K. Freericks and V. Zlatic, 
Condens. Matter Phys. {\bf 9}, 603  (2006).

\bi{Zeb07} M. Zebarjadi, K. Esfarjani, and A. Shakouri,
Appl. Phys. Lett. {\bf 91}, 122104 (2007).

\bi{Chri96} T. Christen and M. B\"uttiker, 
Europhys. Lett. {\bf 35}, 523 (1996); M. B\"uttiker and T. Christen, 
in {\it Mesoscopic Electron Transport}, NATO ASI Series E, Eds. L. Kouwenhoven, G. Schoen,
and L. Sohn, Kluwer (1997).

\bi{Bog99} E.N. Bogachek, A.G. Scherbakov, and U. Landman, 
Phys. Rev. B {\bf 60}, 11678  (1999).

\bi{Cip04} M.A. \c{C}ipilo\u{g}lu, S. Turgut, and M. Tomak, 
Phys. Status Solidi B {\bf 241}, 2575  (2004).

\bi{San12} D. Sanchez and R. Lopez, arXiv:1209.1264.

\bi{caveat} Eq.~(\ref{eq:zteff}) 
has to be modified when time-reversal symmetry is broken by a field
$\phi$ in systems where the thermopower is not symmetric in $\phi$,
$S(\phi) \ne S(-\phi)$. Such systems include Andreev systems, 
see Ref.~[\onlinecite{RobTP}], and
multiterminal devices, see: 
G. Benenti, K. Saito, and G. Casati, Phys. Rev. Lett. {\bf 106}, 230602 (2011).

\bi{But90} P.N. Butcher, 
J. Phys.: Condens. Matter {\bf 2}, 4869 (1990).

\bi{Onsager} Ph. Jacquod, R.S. Whitney, J. Meair, and M. B\"uttiker, Phys. Rev. B {\bf 86}, 155118 (2012).

\bi{thermopower_note} In this work we define the thermopower as $S\equiv(\Delta V/\Delta T)_{I=0}$ whereas in other works it is often defined with an additional minus sign, $S\equiv-(\Delta V/\Delta T)_{I=0}$.

\bi{Pat91} N.K. Patel, J.T. Nicholls, L. Mart\'{i}n-Moreno, M. Pepper, J.E.F. Frost, D.A. Ritchie, and 
G.A.C. Jones, Phys. Rev. B {\bf 44}, 13549 (1991).

\bi{Pro91} C.R. Proetto, Phys. Rev. B {\bf 44}, 9096 (1991).

\bi{Butt90} M. B\"uttiker, Phys. Rev. B {\bf 41}, 7906 (1990).

\bi{Con68} J.N.L. Connor, Mol. Phys. {\bf 15}, 37 (1968).

\bi{Mil68} W.H. Miller, J. Chem. Phys. {\bf 48}, 1651 (1968).

\bi{Ped97} M. H. Pedersen, S. A. van Langen, and M. B\"uttiker, Phys. Rev. B {\bf 57}, 1838 (1997)



\bi{Rob12.2}  R.S. Whitney, arXiv:1211.4737.

\end{thebibliography}
\end{document}